\documentclass[journal]{IEEEtran}
\usepackage{graphicx}

\usepackage{booktabs}

\usepackage{rotating}
\usepackage{tabularx}

\usepackage{tikz}
\newcommand{\rqtwo}[1]{\tikz[baseline={(a.base)}]\node[draw=green!75!black, line width=0.7pt, rounded corners=0.8ex, fill=green!10!white, inner sep=1.5pt,text=black](a){#1};}

\usepackage{xcolor}
\definecolor{graphPurple}{RGB}{170, 0, 255}  

\newcommand{\rqpurple}[1]{%
  \tikz[baseline={(a.base)}]%
    \node[draw=graphPurple!80!black, line width=0.7pt, rounded corners=0.8ex,
          fill=graphPurple!15!white, inner sep=1.5pt, text=black](a){#1};%
}

\newcommand{\rqthree}[1]{\tikz[baseline={(a.base)}]\node[draw=orange!75!black, line width=0.7pt, rounded corners=0.8ex, fill=orange!10!white, inner sep=2pt, text=black](a){#1};}

\usepackage{amssymb}
\usepackage{pifont}
\definecolor{DarkGreen}{rgb}{0.0, 0.5, 0.0}
\newcommand{\greencheck}{{\color{DarkGreen}\checkmark}}
\newcommand{\xmark}{\color{red}\ding{55}}%

\ifCLASSINFOpdf
\else
\fi
\usepackage{amsmath}
\usepackage{amssymb}
\usepackage{algorithm}
\usepackage{algorithmic}

\usepackage{hyperref}
\usepackage{multirow}

\newcommand{\orcid}[1]{\href{https://orcid.org/#1}{\includegraphics[width=0.32cm]{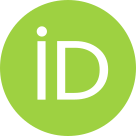}}}

\begin{document}
%
\title{Synthetic generation of online social networks through homophily}
%
%
%

\author{Alejandro~Buitrago~L\'opez\orcid{0009-0002-1606-8766},~\IEEEmembership{}Javier~Pastor-Galindo\orcid{0000-0003-4827-6682}~\IEEEmembership{}and Jos\'e~A.~Ruip\'erez-Valiente\orcid{0000-0002-2304-6365},~\IEEEmembership{Senior~Member,~IEEE}~

\thanks{Alejandro Buitrago L\'opez and Jos\'e~A.~Ruip\'erez-Valiente are in Faculty of Computer Science, University of Murcia, Murcia (Spain).}
\thanks{Javier~Pastor-Galindo is in Computer Systems Engineering Department, Universidad Politecnica de Madrid, Madrid (Spain).}
\thanks{Corresponding author e-mail: \href{mailto:javier.pastor.galindo@upm.es}{javier.pastor.galindo@upm.es} }
\thanks{Manuscript received XXXXX XX, 2025; revised XXXXX XX, 2025.}}

%
%

\markboth{IEEE TRANSACTIONS ON COMPUTATIONAL SOCIAL SYSTEMS,~Vol.~XX, No.~X, XXXXXX~2025}%
{Shell \MakeLowercase{\textit{et al.}}: Bare Demo of IEEEtran.cls for IEEE Journals}
%



\maketitle


\begin{abstract}
Online social networks (OSNs) have become increasingly relevant for studying social behavior and information diffusion. Nevertheless, they are limited by restricted access to real OSN data due to privacy, legal, and platform-related constraints. In response, synthetic social networks serve as a viable approach to support controlled experimentation, but current generators reproduce only topology and overlook attribute-driven homophily and semantic realism.

This work proposes a homophily-based algorithm that produces synthetic microblogging social networks such as $\mathbb{X}$. The model creates a social graph for a given number of users, integrating semantic affinity among user attributes, stochastic variation in link formation, triadic closure to foster clustering, and long-range connections to ensure global reachability. A systematic grid search is used to calibrate five hyperparameters (affinity strength, noise, closure probability, distant link probability, and candidate pool size) for reaching five structural values observed in real social networks (density, clustering coefficient, LCC proportion, normalized shortest path, and modularity).

The framework is validated by generating synthetic OSNs at four scales ($10^3$--$10^6$ nodes), and benchmarking them against a real-world Bluesky network comprising 4 million users. Comparative results show that the framework reliably reproduces the structural properties of the real network. Overall, the framework outperforms leading importance-sampling techniques applied to the same baseline. The generated graphs capture topological realism and yield attribute-driven communities that align with sociological expectations, providing a realistic, scalable testbed that liberates social researchers from relying on live digital platforms.
\end{abstract}


%
\IEEEpeerreviewmaketitle

\section{Introduction}
%
%
%
%
\IEEEPARstart{T}{he} emergence of online social networking services over the past decade has revolutionized how people interact, exchange information, and form communities, making them crucial environments for studying modern social dynamics \cite{Acemoglu2011}. Their large-scale digital traces offer opportunities to model and research phenomena such as polarization, virality, and influence propagation \cite{pastorgalindo2025}.

Despite the societal and scientific importance of online social networks (OSNs), researchers face increasing barriers to accessing meaningful data. Most large-scale platforms do not publish follower graphs or detailed interaction traces, and available APIs typically offer rate-limited, incomplete views of the network \cite{10879487}. Ethical and legal restrictions surrounding user privacy further limit the feasibility of collecting or sharing realistic OSN data at scale. Beyond limited data access, running controlled interventions on live platforms is virtually unfeasible: any algorithmic or content-level modification requires platform approval and risks disrupting user experience. This leaves little room for causal experimentation.

As a result, several methodological alternatives have been explored to study social dynamics in OSNs without relying on full access to platform data. These include analytical modeling, agent-based simulations, synthetic data generation, and experimental platforms under controlled conditions. Among them, social network simulation has emerged as a particularly valuable tool, as it enables the exploration of hypothetical scenarios and intervention strategies. However, the validity of such simulations critically depends on the realism of the underlying network structure. This creates a growing demand for synthetic network generators that can emulate both the structural and semantic characteristics of real-world OSNs \cite{10492674}.

Nevertheless, many existing models for synthetic network generation focus on reproducing structural properties such as degree distributions, clustering, or community structure, often relying on predefined topological rules or statistical patterns. Classical approaches such as Erd\H{o}s--R\'{e}nyi \cite{Erdos1959}, small-world networks \cite{watts1998}, and stochastic block models \cite{holland1983} typically define links through simple probabilistic mechanisms without accounting for the individual-level drivers of relationship formation. Most also assume undirected or reciprocal ties, limiting their ability to represent the asymmetric structure of follower-based platforms. While effective in capturing some structural regularities, these models overlook the semantic dimension of social connections, where similarity in interests, beliefs, or attributes guides friendships \cite{chang2025llmsgeneratestructurallyrealistic}.

In this paper, we close that gap by giving every synthetic node a set of individual properties to create links between similar profiles, mirroring not only the global metrics of real platforms but also social dynamics. In particular, we propose a framework to generate synthetic OSNs from scratch based on homophily in semantic space, the principle that individuals are more likely to connect with others who share similar attributes. Each node is initialized with a latent semantic profile, and links are formed based on the similarity between nodes, modulated by parameters that control affinity thresholds, long-range links, and triadic closure. Our framework is lightweight, unsupervised, and does not require training on real network data, enabling the generation of large-scale graphs that integrate both topological and semantic realism. Moreover, the structural properties of the generated networks are evaluated against a real-world follower graph from the Bluesky platform \cite{bluesky2023}. Our results show that the synthetic networks reproduce key topological metrics while maintaining semantic coherence in the resulting connection patterns, like interest-based clusters and topical communities.

The remainder of the paper is organized as follows. Section~\ref{sota} presents related work. Section~\ref{alg} details the proposed framework. Section~\ref{assess} describes the experimental setup and assessment results. Finally, Section~\ref{conclusion} concludes the paper and outlines directions for future work.

\section{Related work}\label{sota}

The creation of artificial social graphs that emulate the connections of real platforms is critical for computational simulations and controlled experiments. Early efforts focused on purely topological generators, which aim to reproduce global social networks' structural properties, such as degree distributions, clustering coefficients, and community structures. Classical models include the  Erd\H{o}s--R\'{e}nyi model \cite{Erdos1959}, which assumes uniform probability of connection between nodes; the Watts–Strogatz model \cite{watts1998}, which captures small-world properties through local rewiring; and the Barab\'{a}si–Albert model \cite{Barabasi99}, which introduces preferential attachment to explain power-law degree distributions. Other structural models, such as the stochastic block model (SBM) \cite{holland1983}, aim to replicate modularity through group-based connection probabilities. While these approaches are valuable for exploring structural regularities, they abstract away the mechanisms underlying social relationship formation and ignore the semantic or behavioral dimensions of users.

To overcome the limitations of structural models, recent research has explored the use of learning-based and rule-based approaches for generating synthetic networks. For instance, Davies et al.~\cite{10879487} compare Graph Neural Networks (GNNs), such as the Gated Recurrent Attention Network (GRAN) \cite{10.5555/3454287.3454670}, with rule-based generators like R-MAT \cite{10.1137/1.9781611972740.43} and BTER \cite{Comandur2011CommunitySA}. The study demonstrates that GNNs effectively capture structural properties, such as clustering and path lengths, but require access to real-world network data and incur high computational costs. Rule-based methods, in turn, lack semantic richness. 

Beyond purely structural or learning approaches, other models have begun incorporating attributes, semantics, or content into the generation process. For example, the model in \cite{5986141} generated ego networks by iteratively adding nodes (ego) based on estimated emotional closeness, following a layered structure that reflects the strength of social relationships (alters). While this captures individual-level semantics, it is limited to ego-centric structures and does not scale to large or diverse network topologies with structural realism and semantic coherence.

In parallel, some researchers have adopted a data-driven approach, constructing synthetic networks based on population statistics and behavioral data. The work by Karra et al.~\cite{8622199}, for example, introduces a method to generate large-scale contact networks using demographic, geographic, and behavioral data. These networks reflect collocation-based interactions and are validated against empirical metrics. While grounded in real-world realism, such models focus on spatial and temporal structure and do not consider semantic affinity among nodes. Complementing this direction, the authors in \cite{Nettleton2016} presented a method for assigning synthetic attributes to existing topologies under configurable homophily constraints. Starting from real or generated data, the method assigns attributes through a controlled propagation process that preserves target distributions while allowing attribute diversity within communities. While this contributes to semantic realism at the attribute level, it assumes a fixed topology and does not model the interplay between affinity and structure during link formation.

More recently, the emergence of large language models (LLMs) has opened new possibilities for generating semantically rich networks through text-based reasoning. For instance, Chang et al.~\cite{chang2025llmsgeneratestructurallyrealistic} proposed prompting LLMs with node attributes and context to generate synthetic social graphs, introducing three prompting strategies, such as Global, Local, and Sequential, that vary in the exposure of network information during generation. Their analysis reveals that LLM-generated networks reproduce demographic homophily across several attributes, though they tend to overemphasize political homophily, potentially introducing unintended bias. While this approach offers a flexible and human-like mechanism for edge formation, it still relies on extensive prompting and text parsing.

In summary, existing approaches prioritize structural realism through rule-based or statistical models or incorporate semantics via supervised learning, ego-centric designs, or language-model-based reasoning. However, they often face limitations in scalability, generalization, or achieving a balance between semantic coherence and structural fidelity. Furthermore, many of these methods have not been validated against empirical OSN data, which limits their applicability in realistic scenarios. As highlighted in Table~\ref{tab:comparison}, our proposal addresses these gaps by introducing a lightweight, unsupervised algorithm that integrates homophily. Unlike previous work, it enables the generation of large-scale social networks that simultaneously capture key topological features and realistic social connections.

\begin{table}[ht]
\caption{Comparison of approaches for synthetic social network generation across four dimensions: structural fidelity (topology), incorporation of attribute-based homophily, reliance on real-world data for training, and scalability to large graphs.}
\label{tab:comparison}
\centering
\renewcommand{\arraystretch}{1.2}
\setlength{\tabcolsep}{4pt}
\begin{tabular}{
    l l 
    >{\centering\arraybackslash}m{0.8cm} 
    >{\centering\arraybackslash}m{0.8cm} 
    >{\centering\arraybackslash}m{1cm} 
    >{\centering\arraybackslash}m{0.8cm}}
\toprule
\textbf{Reference} & \textbf{Type} & 
\rotatebox{40}{Topology} & 
\rotatebox{40}{Homophily} & 
\rotatebox{40}{Data-driven} & 
\rotatebox{40}{Scalable} \\
\midrule
\cite{Erdos1959} & Statistical          & \greencheck & \xmark        & \xmark        & \greencheck \\
\cite{10.1137/1.9781611972740.43} & Rule-based     & \greencheck & \xmark        & \xmark        & \greencheck \\
\cite{10879487}  & GNN-based           & \greencheck & (indirect)    & \greencheck        & \xmark \\
\cite{chang2025llmsgeneratestructurallyrealistic} & LLM-based & \greencheck & \greencheck        & \greencheck        & \xmark \\
\cite{Nettleton2016} & Attribute-injection & \xmark & \greencheck        & \xmark        & \greencheck \\
\textbf{Ours}    & Homophily-based     & \greencheck & \greencheck        & \xmark        & \greencheck \\
\bottomrule
\end{tabular}
\end{table}

\section{Framework proposal}\label{alg}

This section presents the framework for generating synthetic online social networks grounded in semantic homophily. The design targets platforms where connections are directed and non-reciprocal, such as microblogging services $\mathbb{X}$, where users can follow others without requiring mutual consent. This asymmetric structure is associated with differential visibility, hierarchical influence patterns, and polarized community formation \cite{brady2019}, and thus provides a more realistic basis for studying information diffusion and social dynamics. In Section~\ref{assess}, we assess the extent to which this framework reproduces structural and semantic patterns from real-world OSNs.

\begin{figure*}[!t]
\centering
\includegraphics[width=0.8\textwidth]{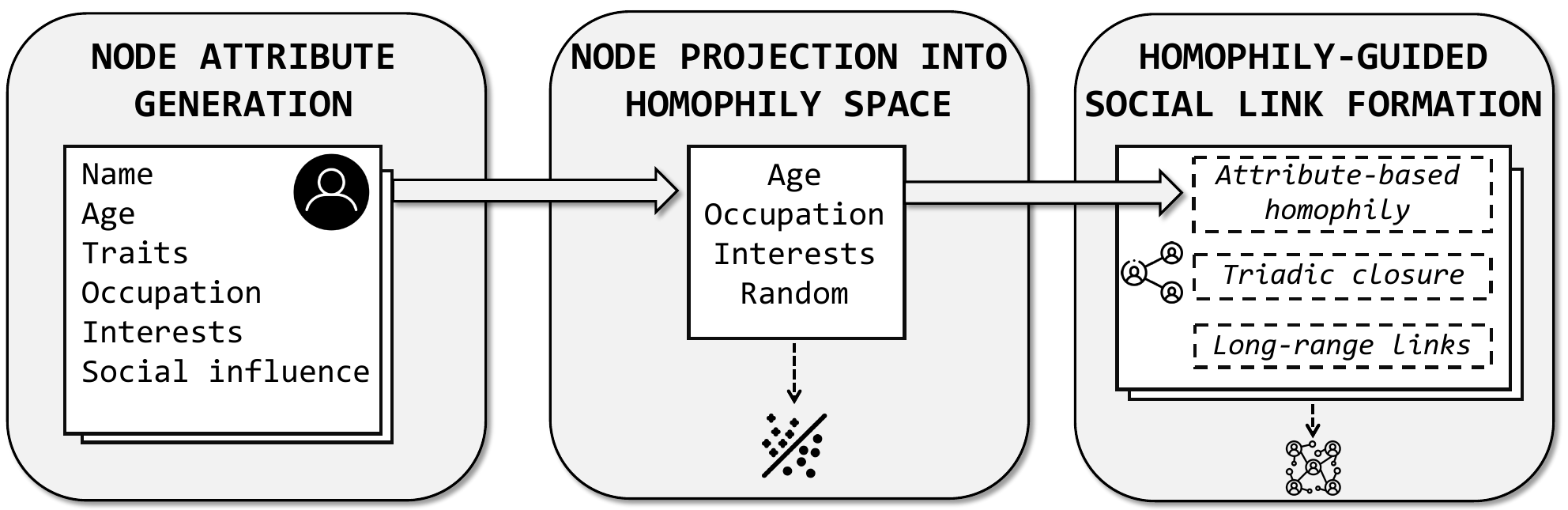}
\caption{Overview of the proposed semantic-homophily-based synthetic network generation.}
\label{fig:algorithm-overview}
\end{figure*}

The generation process unfolds in three phases: first, a set of nodes is initialized, each assigned interests, personality traits, and demographic attributes. Then, nodes are projected into a latent homophily space that encodes semantic similarity. Finally, directed links are formed based on affinity scores modulated by structural parameters controlling local affinity (which promotes connections to nearby similar nodes), long-range links (which enable global connectivity across distant regions of the network), and triadic closure (which reinforces transitive relationships by linking friends-of-friends). The overall workflow is illustrated in Figure~\ref{fig:algorithm-overview}.

\subsection{Node attribute generation}

The generation process begins by assigning each node $n_{i}$ a structured set of attributes that define its semantic profile. According to the article \cite{doi:10.1089/cyber.2018.0670}, each node is assigned a combination of demographic and behavioral attributes, including:

\textit{1) Name:} The name is generated according to the node's gender and is randomly assigned. It serves only as an identifier and does not affect link formation or the resulting network structure.

\textit{2) Age:} Integer that represents the age of the node is sampled based on a custom categorical probability distribution defined over seven age intervals: $[0\text{--}12], [13\text{--}17], [18\text{--}25], [26\text{--}35], [36\text{--}50], [51\text{--}65], [66\text{--}80]$. The distribution parameters are derived from empirical population data reported by Statista for typical online platform users \cite{statista2024age}. For each generation, a discrete probability vector (e.g., $\boldsymbol{p} = [0.01, 0.03, 0.25, 0.30, 0.20, 0.15, 0.06]$) determines the relative frequency of nodes in each age group. The final age is then uniformly drawn within the selected interval~\cite{statista2024age}. This setup ensures demographic realism and stratification aligned with real-world distributions.

\textit{3) Traits:} Five personality traits are modeled using the Big Five personality framework \cite{goldberg1990alternative}, a widely adopted model in psychology and computational social science for characterizing individual behavioral tendencies. This model distinguishes between five major personality traits that describe an individual's personality such as: Neuroticism, Extraversion, Openness (to experience/intellect), Agreeableness, and Conscientiousness. Each trait is operationalized in binary form, representing high or low expression levels (e.g., ``Extraversion +'' or ``Extraversion --''), sampled using predefined probabilities calibrated to reflect typical population variability. In this case, we apply the findings of \cite{doi:10.1073/pnas.2023301118}, assigning a higher probability (75\%) to the positive versions of Extroversion, Neuroticism, and Openness. These traits are later used to modulate each node's social influence score (see point 6).

\textit{4) Occupation:} Due to the lack of publicly available data on occupation distributions in online social networks, we generate this attribute artificially. The occupation of each node is sampled conditionally on the assigned age group. A set of occupation categories is predefined for each age range (e.g., ``Student'' for $13\text{--}25$, ``Retired'' for $66\text{--}80$). One occupation is selected uniformly from the pool associated with the node's age range. This correlation between age and occupation reduces the likelihood of non-realistic profiles and helps maintain semantic consistency across the synthetic population.

\textit{5) Interests:} As with occupation, realistic large-scale data on interest distributions among OSN users is scarce. Therefore, interests are generated synthetically from a predefined list of thematic categories (e.g., ``Technology'', ``Sports''). Each node is assigned up to five interests, randomly sampled uniformly without replacement. This results in heterogeneous interest profiles across the population, which are later used in the affinity calculation stage, where thematic overlap contributes to homophily-based link formation.

\textit{6) Social influence:} Each node's social influence score is initialized by sampling from a global precomputed distribution of influence scores. For this purpose, we employ a distribution that is common in social networks, the power law distribution \cite{10.1145/2567948.2576939}. The raw score is then adjusted based on two factors: (i) high sociability (if ``Extraversion +'' appears among traits), which increases the chance of selecting a top-quintile influence value; and (ii) age, which modulates the influence positively for nodes aged $16$–$39$ and negatively for nodes $40$ and older. The final score is bounded by a maximum threshold and is used later to calculate potential link probability.




Table~\ref{tab:example-nodes} illustrates several examples of synthetic nodes generated through this process. Together, these attributes serve as the semantic foundation for homophily-based link formation in the subsequent stage.

\begin{table}[!t]
\caption{Examples of synthetic nodes generated during attribute initialization.}
\label{tab:example-nodes}
\centering
\footnotesize 
\begin{tabular}{@{}lccccc@{}}
\toprule
\textbf{Name} & \textbf{Age} & \textbf{Occupation} & \textbf{Interests} & \textbf{Traits} & \textbf{Inf.} \\
\midrule
Alice  & 22 & Student  & Tech, Music     & Ext+, Agr+     & 72.5 \\
Bob    & 45 & Engineer & Politics, Sci.  & Con+, Neu+     & 40.2 \\
Emma   & 33 & Designer & Culture, Sport  & Opn+, Ext–     & 55.8 \\
George & 68 & Retired  & Health, History & Neu–, Con–     & 28.4 \\
\bottomrule
\end{tabular}
\end{table}

\subsection{Node projection into homophily space}

Once the semantic attributes have been assigned, each synthetic node is represented by four key components: age, occupation, a set of interests, and a random factor. Each of these attributes is individually transformed into a numerical form. The resulting values are then concatenated into a single vector representation per node, which enables efficient similarity computation to support affinity-based link formation.

To incorporate semantic coherence, categorical attributes such as occupation and interests are not treated as arbitrary labels. Instead, we use precomputed FastText embeddings to compute pairwise cosine similarities between all terms in each domain, as they have shown slightly superior performance in capturing semantic neighborhood structure \cite{Ramadhani_Sena2022}. These similarities are used to perform hierarchical clustering, producing an ordered list in which adjacent entries share high semantic proximity.

From these ordered lists, we construct two dictionaries: \texttt{OCCUPATION\_MAP} and \texttt{INTEREST\_MAP}. Each dictionary maps a semantic identifier to a rank-based ordinal index that reflects its position in the semantically sorted list. These indices are then used as numerical encodings for each categorical term, and subsequently normalized to the $[0,1]$ interval based on the total number of items. Therefore, the assigned position in this sorted list is referred to as the \textit{semantic index}. For example, occupations such as ``Engineer'' and ``Technician,'' which are close in meaning, will be mapped to adjacent indices (e.g., 5 and 6), while semantically distant terms like ``Artist'' or ``Farmer'' will be further apart. Each attribute is encoded as follows:

\begin{itemize}
    \item \textbf{Age} ($A_i$): A scalar in $[0,1]$, obtained by min-max normalization across the age range.
    
    \item \textbf{Occupation} ($O_i$): The occupation label is looked up in \texttt{OCCUPATION\_MAP}, a semantically sorted dictionary of occupations based on precomputed embeddings. The assigned index (i.e., the position in the list) is normalized over the total number of occupations to yield a scalar in $[0,1]$.
    
    \item \textbf{Interests} ($I_i = [I_{i1}, \ldots, I_{ik}]$): Each interest tag is mapped to its semantic index from \texttt{INTEREST\_MAP}, producing a vector of normalized indices. If a user has fewer than $k$ interests, the vector is zero-padded.
    
    \item \textbf{Random component} ($R_i$): A scalar in $[0,1]$ sampled uniformly at random to introduce variability and avoid duplicate embeddings among users with identical attribute combinations.
\end{itemize}

Each node's complete profile is encoded as a vector $x_i \in \mathbb{R}^d$ formed by concatenating all attributes. All scalar components are normalized individually to the $[0,1]$ interval before concatenation. A weight vector $w \in \mathbb{R}^d$ is then applied to modulate the relative influence of each component:  \( \tilde{x}_i = x_i \odot w\) where $\odot$ denotes element-wise multiplication. In our experiments, we use a uniform weighting scheme with $w = [1, 1, \ldots, 1]$, assigning equal importance to all attributes by default. We use uniform weights as a baseline, leaving adaptive tuning of $w$ as future work.

The node projection process is formalized in Algorithm~\ref{alg:homophily-space}, which details how node attributes are transformed into their weighted vector representations for use in affinity computation. The algorithm iterates over each node $n_i \in \mathcal{N}$ and constructs its semantic vector $\tilde{x}_i$ by extracting and normalizing the scalar age $A_i$, mapping the occupation to its semantic index $O_i$ via \texttt{OCCUPATION\_MAP}, and converting each interest in $I_i$ using \texttt{INTEREST\_MAP}. The interest vector is zero-padded to a fixed length $k$ to ensure uniform dimensionality. A random scalar $R_i \sim \mathcal{U}(0,1)$ is appended to introduce controlled variability. All components are concatenated into a vector $x_i$, then modulated by a weight vector $w$ via element-wise multiplication to yield $\tilde{x}_i = x_i \odot w$. The result is the projection matrix \( P \in \mathbb{R}^{N \times d} \), where each row corresponds to a weighted vector \( \tilde{x}_i \) representing node \( n_i \). 

\vspace{0.5em}
\begin{algorithm}[!t]
\caption{Node projection into homophily space}
\label{alg:homophily-space}
\begin{algorithmic}[1]
\REQUIRE List of nodes $\mathcal{N}$, weight vector $w$, semantic maps \texttt{OCCUPATION\_MAP}, \texttt{INTEREST\_MAP}
\ENSURE Matrix $P$ with $|\mathcal{N}|$ rows and $d$ dimensions, where each row corresponds to the weighted semantic vector of a node
\STATE Initialize empty matrix $P$
\STATE Determine $k \gets$ maximum number of interests across nodes
\FORALL{$n_i \in \mathcal{N}$}
    \STATE Extract $A_i \gets$ normalized age of node $i$ in $[0,1]$
    \STATE Retrieve $O_i \gets$ semantic index of occupation, normalized via \texttt{OCCUPATION\_MAP}
    \STATE Initialize interest vector $I_i$ of length $k$ with zeros
    \FORALL{interest $j$ in node $i$'s interests}
        \STATE Map $j$ to semantic index using \texttt{INTEREST\_MAP} and normalize
        \STATE Assign index to next available slot in $I_i$
    \ENDFOR
    \STATE Sample $R_i \sim \mathcal{U}(0,1)$
    \STATE Concatenate all components: $x_i \gets [A_i, O_i, I_i, R_i]$
    \STATE Apply weights: $\tilde{x}_i \gets x_i \odot w$
    \STATE Append $\tilde{x}_i$ to matrix $P$
\ENDFOR
\RETURN $P$
\end{algorithmic}
\end{algorithm}


\subsection{Homophily-guided social link formation}\label{sec:friendships}

The next stage involves constructing the social graph by establishing links between them.

\subsubsection{Principles for link formation}

The network generation process applies a hybrid link formation mechanism over the semantic (homophily) space, also integrating three empirically grounded principles from social network theory \cite{papachristou2024networkformationdynamicsmultillms}: (i) individuals tend to connect with similar others (homophily) \cite{birds_of_a_feather}, (ii) communities emerge through transitive relationships (triadic closure) \cite{10.1145/2499907.2499908}, and (iii) sparse long-range ties enable global connectivity and navigability \cite{DBLP:journals/corr/abs-1111-4503}. This mechanism is additionally conditioned by node-level social influence, which modulates the number of connections each node attempts to establish, leading to heterogeneous degrees across the network.

\paragraph{Local affinity-based links} Each node $n_{i}$ retrieves a set of $k$ nearest neighbors in the homophily space using a KDTree structure, which enables fast sublinear queries. The number $k$ is scaled logarithmically with network size to avoid overly dense neighborhoods:

\begin{equation}
k = \min\left(\log_2(N) \cdot \gamma,\ K_{\max}\right)
\label{eq:k_neighbors}
\end{equation}

where $N$ is the total number of nodes, $\gamma$ is a scaling factor, and $K_{\max}$ caps the maximum local degree. The number of actual links $n$ that node $n_i$ attempts to form is determined by its social influence score, scaled relative to the network size and bounded by $k$. For each neighbor $j$, a link score is computed as:

\begin{equation}
s(n_{i},j) = \alpha \cdot \exp(-d(n_{i},j)) + \beta \cdot \mathcal{U}(0,1)
\label{eq:link_score}
\end{equation}

where \( d(n_i,j) \) is the Euclidean distance between nodes and \( \mathcal{U}(0,1) \) introduces stochastic variation. These scores are passed through a softmax transformation with fixed temperature \( T = 0.5 \), producing a discrete probability distribution over the \( k \) candidates.

A set of \( n \) neighbors is then selected without replacement, meaning each selected node is excluded from further draws in that iteration. The sampling process is directly governed by the softmax-derived probabilities; no additional threshold or Bernoulli test is applied. The selected neighbors are linked to node \( n_i \), forming the initial local connections of the graph.

\paragraph{Triadic closure} To reinforce community structure and simulate social transitivity, a triangle-closing heuristic is applied. If node $n_{i}$ is connected to node $j$, and $j$ to node $k$, a link between $n_{i}$ and $k$ is formed with a probability that adapts to local density:

\begin{equation}
p_{\text{triadic}}(n_{i},k) = \delta \cdot \left(1 + \frac{1}{1 + \exp(-10 \cdot (0.05 - \rho_i))} \right)
\label{eq:triadic_prob}
\end{equation}

where $\rho_i$ is the local connection density of node $n_{i}$, computed as the ratio of its current neighbors to the total nodes. This sigmoid modulation prevents triadic saturation in already dense regions, while promoting triangle formation in sparser areas. This density-aware scaling ensures the mechanism remains effective across different network sizes, preserving clustering without introducing excessive closure in large graphs \cite{10.1145/2499907.2499908}. For each triadic candidate $k$, a Bernoulli trial with probability $p_{\text{triadic}}(n_{i},k)$ determines whether the link is created.

\paragraph{Long-range (exploratory) links} To enhance global connectivity and reduce network diameter \cite{DBLP:journals/corr/abs-1111-4503}, each node evaluates a subset of semantically distant candidates, i.e., users not included among its local neighbors or prior connections. 

The number of candidates $c$ is determined dynamically based on network size, scaling inversely with $\log_2(N)$:

\begin{equation}
c = \frac{\zeta}{\log_2(N)^\theta}
\label{eq:num_candidates}
\end{equation}

where $\zeta$ and $\theta$ are tunable parameters (\texttt{NUM\_CANDIDATES\_SCALE}) that control exploratory capacity as the network grows.

From a larger pre-sample of non-neighbors, $c$ candidates are selected based on descending semantic distance. To reduce degree bias, particularly the over-selection of hubs, each candidate is evaluated using a degree-aware acceptance probability inspired by the Metropolis-Hastings Random Walk (MHRW)~\cite{5462078}:

\begin{equation}
p_{\text{distant}}(i,u) = \frac{\eta}{1 + \log(\deg(u) + 1)}
\label{eq:distant_prob}
\end{equation}

This formulation penalizes highly connected nodes, discouraging hub over-representation while fostering exploration. Combined with the semantic-distance-based preselection, it balances global connectivity with structural diversity. A candidate is ultimately linked if a Bernoulli trial with success probability $p_{\text{distant}}(i,u)$ succeeds.

\subsubsection{Algorithm for the generation of homophily-based social graphs}

The principles previously introduced constitute the theoretical foundation of our algorithm. To translate these mechanisms into a practical and scalable solution for large-scale OSNs, we implement a structured algorithmic procedure that orchestrates their application in a batch-wise manner. This process is formalized in Algorithm~\ref{alg:friendship-generation}, which details how connections are iteratively constructed by combining affinity-driven selection, transitive reinforcement, and exploratory link formation across the semantic space. 

Each node $i$ is embedded in a semantic space derived from a weighted combination of its attributes, resulting in a vector $P_i$ used to compute affinities and distances. For each node $i$, $k$ nearest semantic neighbors are retrieved via KDTree (Eq.~\eqref{eq:k_neighbors}). A subset of $n$ neighbors is selected based on the node's social influence, scaled relative to the total number of nodes and capped by $k$. Link scores $s(i,j)$ are computed as in Eq.~\eqref{eq:link_score}, then normalized into probabilities via a softmax transformation ($T = 0.5$). A total of $n$ neighbors are sampled without replacement and linked to $i$, updating the bidirectional map $\mathcal{H}$.

Triadic closure is then applied: candidates $k$ are drawn from friends-of-friends not in $\mathcal{H}[i]$, and links are added with a Bernoulli trial using $p_{\text{triadic}}(i,k)$ (Eq.~\eqref{eq:triadic_prob}), which adapts to local density. In large networks ($N \geq 10^5$), an additional triadic closure heuristic may be optionally applied to reinforce community connectivity beyond immediate friends-of-friends.

Finally, long-range candidates $u$ are sampled by descending semantic distance, and the top-$c$ are retained (Eq.~\eqref{eq:num_candidates}). Each candidate is linked with probability $p_{\text{distant}}(i,u)$ (Eq.~\eqref{eq:distant_prob}), modulated by their degree. Each confirmed exploratory link is added to both $\mathcal{F}$ and $\mathcal{H}$.

\vspace{0.5em}
\begin{algorithm}[!t]
\caption{Friendship generation via homophily, closure, and exploration}
\label{alg:friendship-generation}
\begin{algorithmic}[1]
\REQUIRE List of users $\mathcal{N}$; semantic vectors $P$; KDTree $T$; parameters $\alpha$, $\beta$, $\delta$, $\eta$, $\zeta$; neighborhood map $\mathcal{H}$
\ENSURE Edge list $\mathcal{F}$ (global edge list); and updated neighborhood map $\mathcal{H}[i]$ for each user
\FORALL{user $i \in \mathcal{N}$}
    \STATE Retrieve $k$ nearest neighbors of $i$ using $T$ \textit{(Eq.~\eqref{eq:k_neighbors})}
    \STATE Determine $n$ as a function of node $i$'s social influence and $N$
    \STATE Compute $d(i,j)$ as Euclidean distance in semantic space
    \STATE Compute logits $l(j) \gets \alpha \cdot (-d(i,j)) + \beta \cdot \mathcal{U}(0,1)$ \textit{(Eq.~\eqref{eq:link_score})}
    \STATE Compute probabilities via softmax: $p(i,j) \propto \exp(l(j)/T) \quad \text{with } T = 0.5$
    \STATE Select $n$ neighbors from top-$k$ candidates, sampling without replacement using $p(i,j)$
    \FORALL{selected neighbors $j$}
        \STATE Add link $(i,j)$ to $\mathcal{F}$ and update $\mathcal{H}$ by adding $j$ to $\mathcal{H}[i]$ and $i$ to $\mathcal{H}[j]$
    \ENDFOR

    \STATE Identify triadic candidates $k$ from friends-of-friends not in $\mathcal{H}[i]$
    \STATE Compute local density $\rho_i \gets |\mathcal{H}[i]| / N$
    \STATE Set $p_{\text{triadic}}(i,k)$ according to Eq.~\eqref{eq:triadic_prob}
    \FORALL{triadic candidates $k$}
        \IF{$\text{random}(0,1) < \delta_i$}
            \STATE Add link $(i,k)$ to $\mathcal{F}$ and update $\mathcal{H}$
        \ENDIF
    \ENDFOR

    \STATE Pre-sample sorted distant candidates $u \notin \mathcal{H}[i]$ by descending semantic distance
    \STATE Select top-$c$ distant candidates according to Eq.~\eqref{eq:num_candidates}
    \FORALL{candidate $u$}
        \STATE Compute $p_{\text{distant}}(i,u) \gets \eta / (1 + \log(\deg(u) + 1))$ \textit{(Eq.~\eqref{eq:distant_prob})}
        \IF{$\text{random}(0,1) < p_{\text{distant}}(i,u)$}
            \STATE Add link $(i,u)$ to $\mathcal{F}$ and update $\mathcal{H}$
        \ENDIF
    \ENDFOR
\ENDFOR
\RETURN $\mathcal{F}$
\end{algorithmic}
\end{algorithm}

\subsubsection{Controllable hyperparameters and semantic weights}

The link formation mechanism described above can be parametrized to shape the structural and semantic properties of the generated network. By adjusting the following hyperparameters, networks with different topological signatures can be generated, ranging from highly clustered graphs to flatter structures with lower modularity. Crucially, several of these parameters are not fixed values, but rather adaptive functions of the network size \( N \). These functions are calibrated to preserve structural realism across scales, ensuring that properties such as modularity, clustering, and average shortest path remain within empirically plausible ranges:

(1) \texttt{weights} ($w$): Fixed user-defined vector of attribute-level weights applied during projection into homophily space. By tuning the relative importance of components (e.g., interests, age, occupation), one can influence the role each plays in affinity computation.

(2) \texttt{CONN\_EXP\_WEIGHT} ($\alpha$): Governs the weight of semantic affinity in local link formation. Larger values increase the probability of connecting to nearby nodes in the homophily space. Typically fixed, but may be tuned manually.
    
(3) \texttt{CONN\_RAND\_WEIGHT} ($\beta$): Introduces randomness in link scores. Higher values soften strict homophily, increasing link diversity and degree variance. Also typically fixed.

(4) \texttt{TRIADIC\_PROB} (\( \delta(N) \)): Probability of forming transitive links (triadic closure), adaptively scaled with network size. Larger values result in higher clustering coefficients and more cohesive communities. It is computed as:
    \[
    \delta(N) = \min\left(\delta_{\text{cap}},\ \delta_0 + \frac{\lambda}{(\log_2 N)^{\mu}}\right)
    \]
    where \( \delta_0 \) (\texttt{TRIADIC\_PROB\_BASE}) is the base probability, \( \lambda \) (\texttt{TRIADIC\_PROB\_SCALE}) is a scaling factor, and \( \delta_{\text{cap}} \) (\texttt{TRIADIC\_PROB\_CAP}) limits its upper bound.
    
    In small graphs (\( N < 10^4 \)), a suppression factor is applied: \( \delta(N) \gets \delta(N) \cdot 0.35 \) to avoid artificial cliques and reinforced in large ones to preserve clustering. In large graphs (\( N \geq 10^5 \)), a lower bound is enforced: \( \delta(N) \gets \max(\delta(N),\ 0.22 + 0.02 \cdot \log_{10} N) \).

(5) \texttt{Y\_DISTANT\_PROB} (\( \eta(N) \)): Probability of forming links with semantically distant nodes. This parameter enhances global reachability and reduces average path length. It is computed as:
   \[
    \eta(N) = \min\left(\eta_{\text{cap}},\ \eta_0 + \frac{\kappa}{(\log_2 N)^{0.75}}\right)
    \]
    where \( \eta_0 \) (\texttt{Y\_DISTANT\_PROB\_BASE}) is the base distant-link probability, \( \kappa \) (\texttt{Y\_DISTANT\_PROB\_SCALE}) its scaling factor, and \( \eta_{\text{cap}} \) (\texttt{Y\_DISTANT\_PROB\_CAP}), which is computed as \( \eta_{\text{cap}} = \eta_0 + \kappa \) by default, though it can be independently configured, is the upper limit.
    
    For small graphs (\( N < 10^4 \)), the value is suppressed: \( \eta(N) \gets \eta(N) \cdot 0.05 \) to prevent over-bridging. For large graphs (\( N \geq 10^5 \)), a minimum floor is applied: \( \eta(N) \gets \max(\eta(N),\ \eta_0 + 0.02 \cdot \log_{10} N) \) to maintain low diameter and strong connectivity.

(6) \texttt{NUM\_CANDIDATES} (\( c(N) \)): Size of the distant candidate pool:
   \[
    c(N) = \max\left(1,\ \left\lfloor \frac{\zeta}{(\log_2 N)^{\theta}} \right\rfloor \right)
    \]
    where \( \zeta \) (\texttt{NUM\_CANDIDATES\_SCALE}) is the candidate scaling factor and \( \theta = 0.6 \). This value is further suppressed for small graphs (\( c \gets c \cdot 0.25 \) if \( N < 10^4 \)), and bounded from below in large graphs (\( c \gets \max(c, 6) \) if \( N \geq 10^5 \)) to ensure exploratory diversity without overwhelming the local structure.

In the absence of such scaling functions, small graphs would exhibit unrealistic over-connectivity—resulting in densely saturated or near-clique topologies—while large graphs would suffer from sparse connectivity, diminished clustering, and limited navigability. These adjustments are grounded in empirical observations and theoretical constraints, allowing the generator to preserve key topological metrics, such as clustering, modularity, and average shortest path, across several orders of magnitude in network size \cite{chang2025llmsgeneratestructurallyrealistic, 10.1145/1150402.1150479}. 

These parameters are not optimized a priori. Instead, in Section~\ref{assess}, we empirically explore their effects by evaluating multiple configurations and comparing the resulting structural metrics to those of a real-world OSN.

\section{Assessment of synthetic social networks}\label{assess}

To evaluate the effectiveness of the proposed generative framework, we conduct a systematic validation of the synthetic networks by comparing them against a real-world benchmark. This assessment is not part of the generation pipeline itself, but rather a separate methodological component designed to quantify structural realism. Our evaluation strategy involves: (i) defining reference network metrics for structural and semantic comparison, (ii) performing a grid search over key hyperparameters to calibrate the generator toward real-world values, and (iii) benchmarking the best resulting configuration against alternative synthetic baselines, including subsampling methods.

\subsection{Social network characterization metrics}
\label{sec:metrics}

Before analyzing and comparing the properties of social networks, a set of standard metrics of OSNs is established. These variables capture local and global properties relevant to network cohesion, navigability, and community structure \cite{chang2025llmsgeneratestructurallyrealistic}:

(1) \textbf{Density:} Fraction of realized edges (E) over all possible edges within nodes (N). Indicates how interconnected the network is. Sparse connectivity is a characteristic of large OSNs \cite{WONG200699}.

(2) \textbf{Average clustering coefficient:} Measures the tendency of nodes to form closed triads, indicative of local cohesion. For each node $i$, $E_i$ is the number of edges between its neighbors, and $k_i$ is its degree. Social networks are known to exhibit clustering, where one's friends are likely to be friends with each other \cite{10.1145/2499907.2499908}.

(3) \textbf{Large connected component (LCC) proportion:} Social networks are known to be well connected \cite{DBLP:journals/corr/abs-1111-4503}, with the vast majority (over 99\%) of the nodes in the LCC, i.e., the largest subgraph where all nodes within the subgraph are reachable by each other. This metric helps assess the network's connectivity.

(4) \textbf{Normalized average shortest path length:} Social networks are not only well-connected, meaning nodes can reach each other, but they can also reach each other in relatively short paths \cite{watts1998}. This metric helps measure the efficiency of information or influence spread within the network, as it scales with the logarithm of the number of nodes in Erd\H{o}s--R\'{e}nyi graphs \cite{Erdos1959}. Let $\mathcal{P}$ be the set of all node pairs with valid paths, and $\text{SP}(i,j)$ the shortest path length between nodes $i$ and $j$.

(5) \textbf{Modularity:} Social networks exhibit strong community structure, with more edges within communities and fewer edges across communities \cite{Newman2004}. This metric measures the quality of community structure, computed via the Louvain method \cite{Blondel_2008}. High modularity indicates well-defined group separation.

\subsection{Hyperparameter tuning}

Having defined the referencing variables to evaluate a social network, the proposed algorithm is empirically calibrated to approximate the structural properties of real OSNs. The reference for optimization is the follower graph from the decentralized Bluesky platform~\cite{failla_2024_14258401}, comprising over 4M users (nodes) and $144$ million friendships (directed edges). Bluesky, like $\mathbb{X}$ (formerly Twitter) or Mastodon, offers an asymmetric social structure and algorithmic feeds, aligning well with the homophily-based link formation modeled by the proposed algorithm.


The calibration process involves aligning the structural metrics of the generated network with those observed in Bluesky through a grid search over eight key hyperparameters. Semantic affinity in local link formation is governed by $\alpha$. At the same time, stochastic variation is modulated by $\beta$. Triadic closure is controlled by three parameters: the base probability $\delta_0$, scaling factor $\lambda$, and upper cap $\delta_{\text{cap}}$. Long-range connectivity is shaped by the base distant-link probability $\eta_0$, and its scaling $\kappa$. Finally, $\zeta$ controls the size of the exploratory candidate pool. These parameters reflect theoretical principles such as homophily, randomness, clustering, and navigability, and influence emergent topological properties including clustering coefficient, modularity, and average shortest path~\cite{birds_of_a_feather,watts1998}.

Numerical ranges for each hyperparameter were selected based on theoretical foundations, prior empirical evidence, and targeted sensitivity analyses. $\alpha$ was varied in \{0.0, 0.1, 0.16, 0.25\} to capture varying levels of homophily strength~\cite{doi:10.1073/pnas.2023301118}. The stochastic factor $\beta$ explored values in \{0.0, 0.06, 0.1\}, consistent with observed randomness in OSNs~\cite{DBLP:journals/corr/abs-1111-4503}. Triadic closure base probabilities $\delta_0$ were selected from \{0.0, 0.10, 0.20, 0.25\}, with scaling factors $\lambda$ in \{0.5, 1.5, 2.5, 3.5\} and caps $\delta_{\text{cap}}$ in \{0.40, 0.42, 0.44\}, supporting variation in clustering intensity~\cite{10.1145/2499907.2499908}. For long-range links, $\eta_0$ was tested over \{0.0, 0.02, 0.03, 0.04, 0.05\}, and $\kappa$ over \{0.01, 0.05, 0.1\}, to balance modularity with small-world effects~\cite{watts1998}. The distant candidate pool size $\zeta$ was drawn from \{8, 12, 24, 36\}, controlling the breadth of exploratory attempts while preserving local structure~\cite{papachristou2024networkformationdynamicsmultillms}.

In total, $34,561$ configurations were tested, resulting from the Cartesian product of all parameter combinations. Each configuration was evaluated by generating a synthetic network of size $N=10^3$ (allowing scalable iteration while preserving informative structural patterns~\cite{10.1145/2499907.2499908}) and calculating the five structural metrics described in Section~\ref{sec:metrics}. These values are compared with the full Bluesky network metrics (density = $8.6 \times 10^{-6}$, average clustering coefficient = $0.262$, LCC proportion = $1.0$, average shortest path length = $0.230$, and modularity = $0.85$) through the Normalized Euclidean Distance (NED).

Table~\ref{tab:top-configs} shows the five configurations with the lowest NED to the real network. High structural fidelity was linked to strong triadic closure (\( \lambda \geq 2.5 \), \( \delta_0 \in [0.20, 0.25] \)), relaxed caps (\( \delta_{\text{cap}} \geq 0.40 \)), and semantic affinity \( \alpha = 0.16 \). Best results also featured low randomness (\( \beta \leq 0.06 \)), long-range link probabilities \( \eta_0 \in [0.04, 0.05] \) with \( \kappa \geq 0.05 \), and large candidate pools (\( \zeta = 36.0 \)), highlighting the value of exploratory reach.

\begin{table}[!t]
\centering
\caption{Top five hyperparameter configurations ranked by structural fidelity to the Bluesky graph ($N=10^3$). Lower NED indicates higher realism.}
\label{tab:top-configs}
\begin{tabular}{cccccccc|c}
\toprule
\multicolumn{8}{c|}{\textbf{Algorithm hyperparameters}} & \textbf{NED} \\
$\eta_0$ & $\kappa$ & $\delta_0$ & $\lambda$ & $\delta_{\text{cap}}$ & $\zeta$ & $\alpha$ & $\beta$ & \\
\midrule
0.05 & 0.10 & 0.20 & 3.5 & 0.42 & 36.0 & 0.16 & 0.06 & 0.852 \\
0.02 & 0.01 & 0.25 & 0.5 & 0.40 & 36.0  & 0.1 & 0.06 & 0,899 \\
0.05 & 0.05 & 0.20 & 2.5 & 0.40 & 24.0 & 0.16 & 0.00 & 0,900 \\
0.04 & 0.10 & 0.20 & 2.5 & 0.40 & 8.0  & 0.16 & 0.10 & 0.901 \\
0.05 & 0.05 & 0.25 & 2.5 & 0.40 & 36.0 & 0.16 & 0.00 & 0.904 \\
\bottomrule
\end{tabular}
\end{table}

The best-performing configuration is selected as the default setup for the comparative evaluation in Section~\ref{assess:comparative}, where it is benchmarked against state-of-the-art sampling methods.

\subsection{Comparative evaluation of synthetic networks with real OSN samples}\label{assess:comparative}

Having identified the most realistic parameter configuration via grid search, we assess the realism of the generated synthetic networks at four scales, from $10^3$ to $10^6$ nodes. Under the optimized setup, synthetic networks are compared against subgraphs of the same size. The latter are extracted from the Bluesky graph using structure-preserving sampling techniques: Forest Fire, Random Walk, PageRank Node, and Metropolis Hastings Random Walk, previously employed to validate structural fidelity~\cite{5462078,6027868}. Additionally, Random Node sampling and the full Bluesky graph are included as lower (worst case) and upper (best case) baselines, respectively.

\subsubsection{Characterization metrics of synthetic networks}

\begin{figure*}[!t]
\centering
\includegraphics[width=0.95\textwidth]{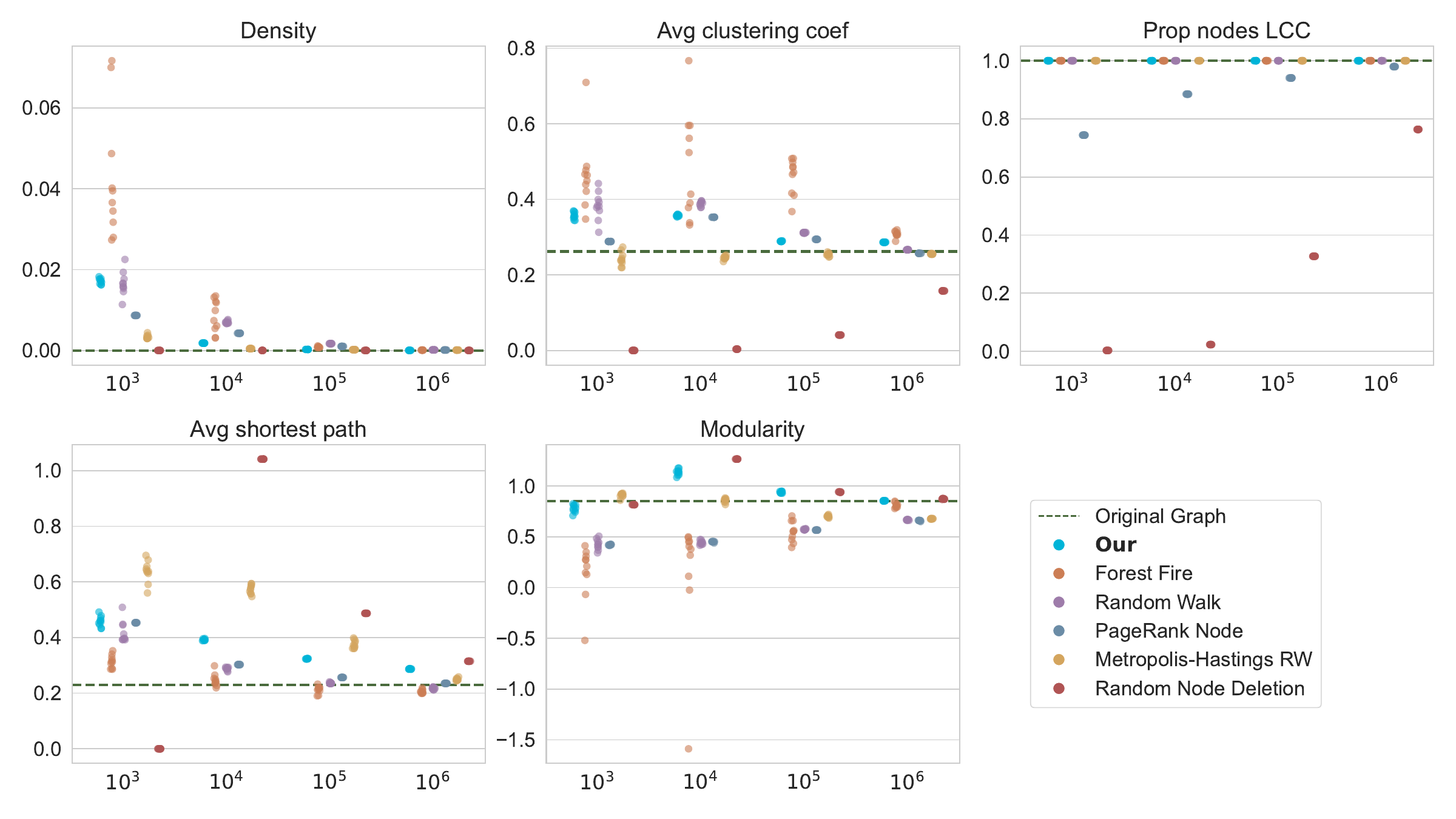}
\caption{
Comparison of network structure across our synthetic networks and real-world sampled networks at four scales ($10^3$ to $10^6$ nodes). Each subplot shows one structural metric. The dashed horizontal line represents the corresponding value in the real-world Bluesky graph.
}
\label{fig:metric-comparison}
\end{figure*}

Figure~\ref{fig:metric-comparison} summarizes the results for the five characterization metrics introduced in Section~\ref{sec:metrics}. Each panel presents the mean and standard deviation across ten runs, alongside individual data points. The horizontal dashed line represents the value observed in the full Bluesky network (real-world referencing value).

Across most metrics and scales, the synthetic networks generated by our method (in blue) closely mimic the Bluesky network and exhibit low variance, demonstrating both accuracy and stability. In terms of average clustering coefficient and modularity, our synthetic networks consistently match or exceed the fidelity of sampled real ones. The average shortest path length remains close to the reference value, suggesting that long-range links are sufficient to maintain navigability. Similarly, the proportion of nodes in the LCC remains consistently high and comparable to that of sampled graphs.

These results demonstrate that our proposal does not merely approximate global graph statistics but produces structurally coherent and resilient networks across a range of sizes, achieving high realism that even sampling techniques do not reach.

\subsubsection{Realism of synthetic networks}

To synthesize the multi-dimensional results into a unified indicator of structural realism, a complementary analysis is performed using the NED between each generated or sampled network and the full Bluesky graph. This measure aggregates deviations across the five structural metrics defined in Section~\ref{sec:metrics}, after min-max normalization across all methods and sizes.

\begin{figure*}[!ht]
\centering
\includegraphics[width=0.91\textwidth]{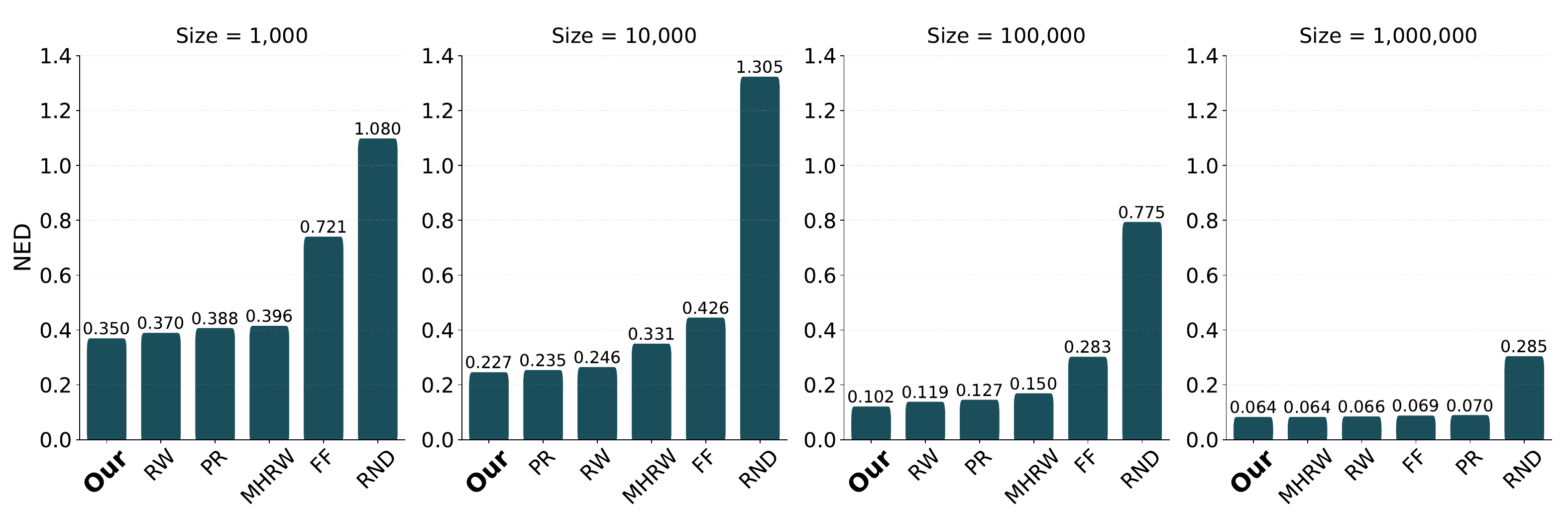}
\caption{NED to the full Bluesky graph at four scales ($10^3$ to $10^6$ nodes). Lower values indicate higher structural realism. Values are averaged over 10 runs per method and scale.}
\label{fig:distance-summary}
\end{figure*}

Figure~\ref{fig:distance-summary} displays these aggregated distances for our synthetic networks and sampled networks at four size scales. Each subplot reports the average distance across ten instances per network, with lower values indicating closer structural alignment with the reference network. 

As illustrated in the figure, the distance to realistic values always decreases when the size increases, indicating that larger synthetic and sampled graphs reproduce the real-world network structural patterns more faithfully. Moreover, our proposed algorithm consistently achieves the lowest or near-lowest distance at every scale, surpassing all sampling baselines not only at small and medium sizes ($10^3$–$10^4$), but also at the largest scale ($10^5$–$10^6$).

\subsection{Homophily evaluation of synthetic social networks}

\begin{figure}[!ht]
\centering
\includegraphics[width=0.75\columnwidth]{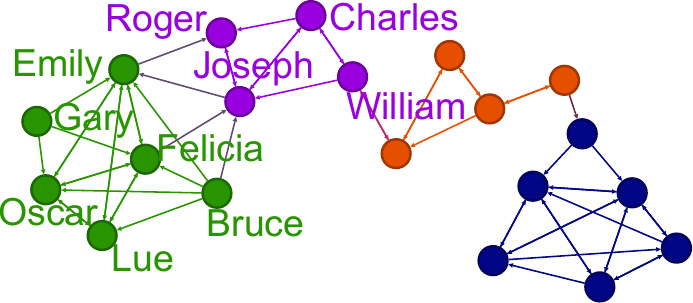}
\caption{Synthetic social network graph with color-coded communities. Nodes represent users, edges represent following relationships, and node labels indicate user names. Community detection via modularity reveals structurally cohesive groups with internal attribute similarity (homophily).}
\label{fig:social-graph}
\end{figure}

In addition to structural metrics, it is important to evaluate whether the generated social connections are semantically coherent. That is, we assess whether nodes tend to connect with others who have similar attributes, reflecting realistic patterns of homophily observed in real online social networks. Figure~\ref{fig:social-graph} provides a qualitative example of a synthetic network of $N=30$ nodes generated with the best-performing configuration. Nodes are colored according to the communities detected using the Louvain algorithm, a widely adopted method for community detection \cite{Blondel_2008}. Notably, the structure reveals four clearly differentiated communities, each densely connected internally and sparsely connected across boundaries, consistent with modular topologies observed in real OSNs.

To assess whether the detected communities also reflect semantic affinity, two representative groups in green and purple are analyzed in detail in Table~\ref{tab:communities_2_3_homophily}. These groups exhibit a strong alignment between structural connectivity and attribute-level similarity. For instance, the green community comprises young users employed in service roles (e.g., waiter, employee), sharing interests such as basketball, crafts, and painting, as well as cooperative or emotionally expressive traits. In contrast, the purple group consists of older professionals (e.g., chemist, psychologist), connected through shared educational background, introverted or dependable traits, and creative or intellectual hobbies like photography and classical music.

The remaining communities, orange and blue ones, exhibit similarly consistent patterns. The orange cluster gathers mid-aged users with moderate education and a blend of emotional and aesthetic traits, while the blue group contains a highly interconnected subgroup with shared occupational paths and overlapping psychological profiles. Importantly, cross-cluster ties are minimal and, when present, often bridge semantically adjacent profiles (e.g., shared interests or intermediary neighbors), reinforcing the plausibility of the generated link structure.

\begin{table*}
\centering
\caption{Detailed attributes and following relationships of users in green and purple, illustrating structural and attribute-based homophily through shared occupations, interests, and psychological traits.}
\label{tab:communities_2_3_homophily}
\begin{tabular}{p{0.5cm} r p{0.6cm} l l p{3cm} p{3.5cm} p{3.5cm}}
\toprule
Name & Age & Gender & Occupation & Qualification & Interests & Traits & Relations \\
\midrule
\rqtwo{Lue}     & 20 & Female & Waiter      & PhD                    & Basketball             & Expressiveness, Emotionality     & \rqtwo{Oscar}, \rqtwo{Felicia}, \rqtwo{Emily} \\
\rqtwo{Oscar}   & 19 & Male   & Employee    & Primary School         & Painting               & Sociable                        & \rqtwo{Lue}, \rqtwo{Felicia}, \rqtwo{Emily} \\
\rqtwo{Gary}    & 18 & Male   & Employee    & Primary School         & Cooking                & Independence, Discipline         & \rqtwo{Oscar}, \rqtwo{Felicia}, \rqtwo{Emily} \\
\rqtwo{Bruce}   & 20 & Male   & Waiter      & Middle School          & Contemporary Music     & Independence, Emotionality       & \rqtwo{Lue}, \rqtwo{Oscar}, \rqpurple{Joseph}, \rqtwo{Felicia}, \rqtwo{Emily} \\
\rqtwo{Felicia} & 20 & Female & Waiter      & High School Diploma    & Crafts                 & Cooperation, Passivity  & \rqtwo{Lue}, \rqtwo{Oscar}, \rqpurple{Joseph}, \rqtwo{Emily} \\
\rqtwo{Emily}   & 22 & Female & Builder     & Professional Certificate & Basketball           & Cooperation, Lethargy            & \rqtwo{Lue}, \rqtwo{Oscar}, \rqpurple{Roger}, \rqtwo{Felicia} \\
\midrule
\rqpurple{Charles} & 29 & Male & Chemist     & Master's Degree        & Board Games            & Dependability                    &\rqpurple{ William}, \rqpurple{Joseph}, \rqpurple{Roger} \\
\rqpurple{William} & 33 & Male & Psychologist & Associate Degree      & Photography            & Emotionality, Optimism           & \rqthree{Nicole}, \rqpurple{Charles}, \rqpurple{Joseph} \\
\rqpurple{Joseph}  & 27 & Male & Police      & High School Diploma    & Nature Photography     & Independence, Spontaneity        & \rqpurple{Charles}, \rqpurple{Roger}, \rqtwo{Emily} \\
\rqpurple{Roger}   & 27 & Male & Trucker     & Master's Degree        & Classical Music        & Emotionality                     & \rqpurple{Joseph} \\
\bottomrule
\end{tabular}
\end{table*}



\subsection{Performance evaluation of the generation of synthetic social networks}

\begin{figure}[ht!]
  \centering
  \includegraphics[width=0.8\columnwidth]{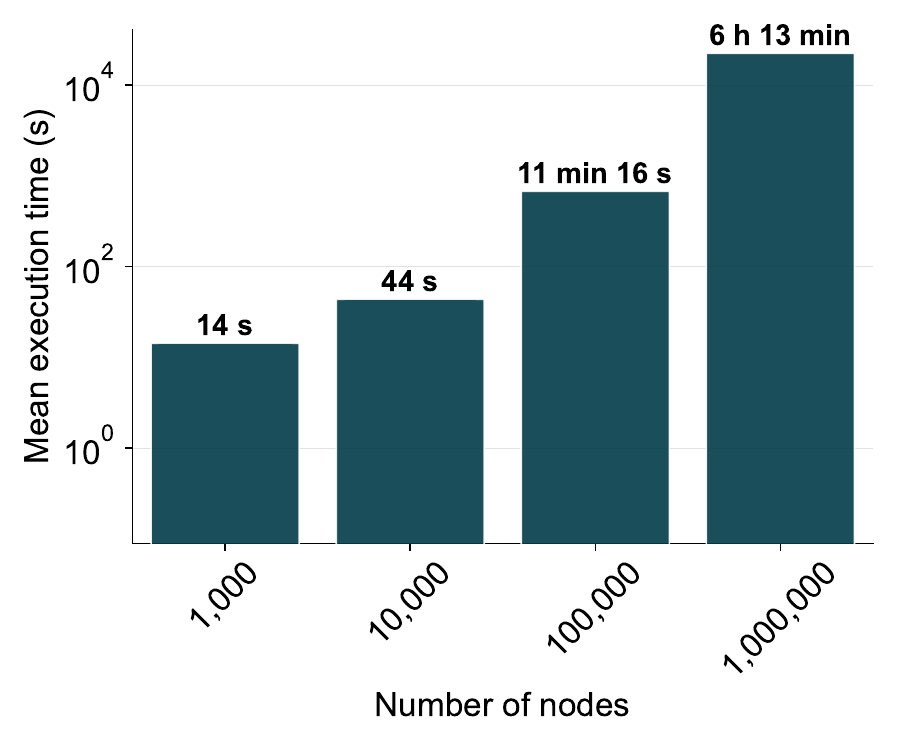}
  \caption{Mean execution time (over 10 generations) of the proposed framework as a function of the number of nodes.}
  \label{fig:scaling}
\end{figure}

To assess the scalability of our approach, we measured the mean execution time across 10 independent generations for increasing network sizes on a 10-core, 100\,GB RAM Linux server. As shown in Figure~\ref{fig:scaling}, the framework completes in 14 seconds for 1,000 nodes, 44 seconds for 10,000, 11.3 minutes for 100,000, and 6.2 hours for 1\,million nodes. The growth remains tractable up to mid-sized networks, with execution time increasing subquadratically due to affinity-based neighbor queries and link sampling. To the best of our knowledge, there are no prior studies that systematically report execution times for synthetic OSN generation. Our results, therefore, offer a useful baseline for future implementations and performance comparisons.

\section{Conclusion}\label{conclusion}

Studying and intervening in OSNs presents major challenges due to their scale, complexity, and restricted access to complete data. Synthetic network generators offer a promising alternative, enabling controlled experimentation and counterfactual analysis. However, realism, both structural and semantic, remains a key requirement for such simulations to reflect real-world behavior meaningfully.

This work proposed a parametrized framework for generating homophily-based synthetic social networks, combining affinity-driven link formation, triadic closure, and long-range exploration with semantically rich node representations. Nodes are endowed with a set of demographic and behavioral attributes, including age, occupation, interests, and personality traits. The framework is transparent and explainable by design, with interpretable hyperparameters that can be calibrated to emulate different topologies and social tendencies.

The framework is evaluated by extensive experiments across four network sizes from $10^3$ to $10^6$ nodes, demonstrating that generated synthetic networks highly approximate the network structure of a real-world Bluesky network and outperform associated sampled subgraphs. Comparative experiments highlight the critical role of network size in determining fidelity, with deviations diminishing at larger scales. Moreover, the framework produces cohesive, semantically plausible communities aligned with observed patterns of homophilic social formation. Quantitatively, the normalized error between generated and real networks remains low across key metrics, suggesting the method effectively captures the underlying structure it seeks to emulate.

Future work will focus on incorporating attribute correlations, temporal dynamics, and emergent phenomena such as polarization or echo chambers. The framework's modularity and realism offer a foundation for simulating interventions, algorithmic policies, or social behaviors under conditions that preserve key empirical features of OSNs.

\section*{Acknowledgment}

This work has been partially funded by (a) the strategic project CDL-TALENTUM from the Spanish National Institute of Cybersecurity (INCIBE), the Recovery, Transformation, and Resilience Plan, Next Generation EU, (b) by the University of Murcia by FPU contract, and (c) by a ``Juan de la Cierva'' Postdoctoral Fellowship (JDC2023-051658-I) funded by the i) Spanish Ministry of Science, Innovation and Universities (MCIU), ii) by the Spanish State Research Agency (AEI/10.13039/501100011033) and iii) by the European Social Fund Plus (FSE+). The authors would also like to thank Antonio Guill\'{e}n for his valuable contributions to the development of the node generation algorithm.

\ifCLASSOPTIONcaptionsoff
  \newpage
\fi



%
\bibliographystyle{IEEEtran}
\bibliography{bibtex/bib/bibliography.bib}

%

\begin{IEEEbiography}[{\includegraphics[width=1in,height=1.25in,clip,keepaspectratio, angle=-90]{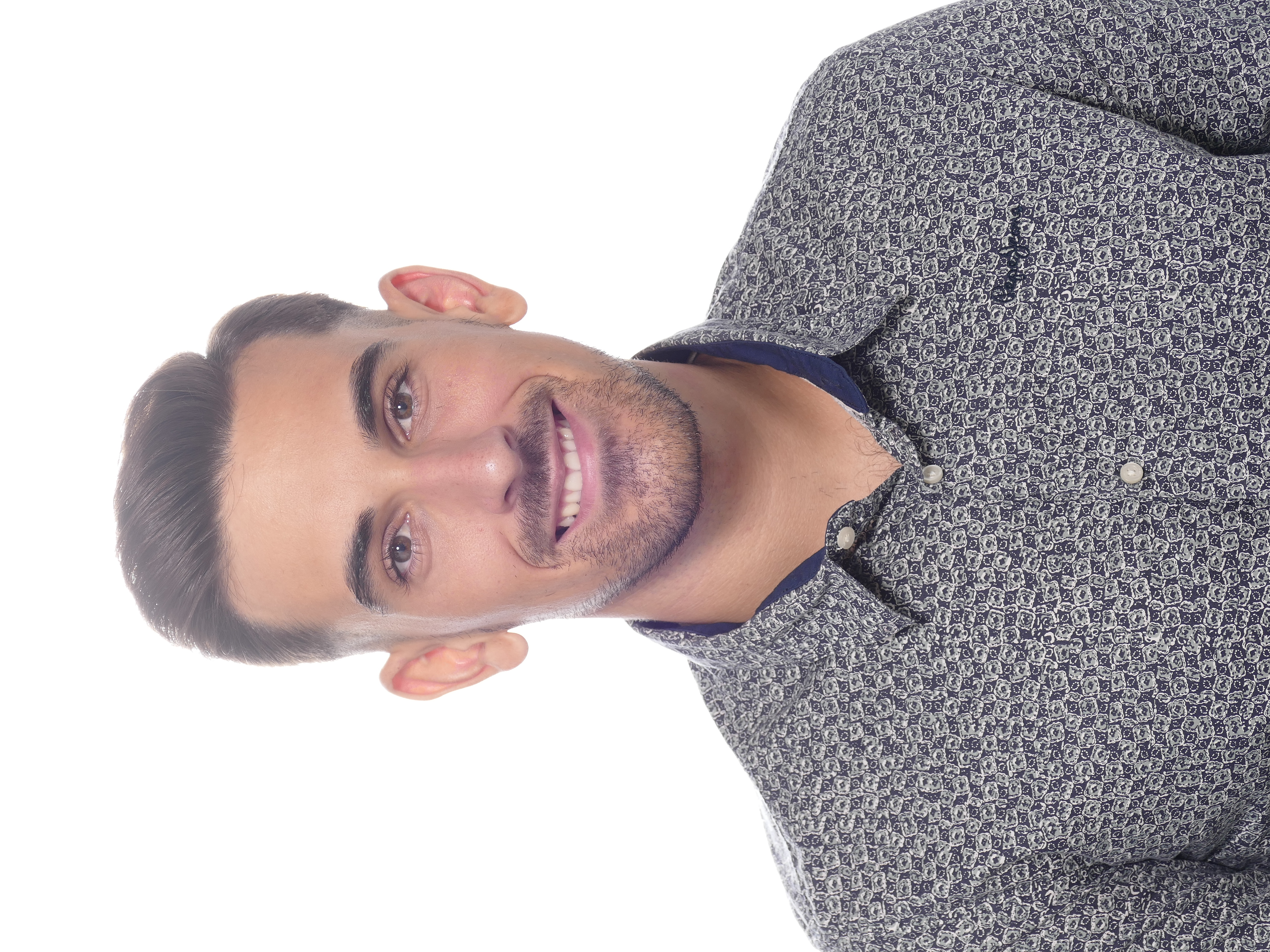}}]{Alejandro Buitrago L\'{o}pez}
is working towards a Ph.D. in Computer Science at the University of Murcia, Spain. He obtained a B.Sc. Degree with a focus on software engineering and a M.Sc. in Big Data. He is a member of the CyberDataLab at the University of Murcia, and his research interests include data science, disinformation, and cybersecurity.
\end{IEEEbiography}

\begin{IEEEbiography}[{\includegraphics[width=1in,height=1.25in,clip,keepaspectratio]{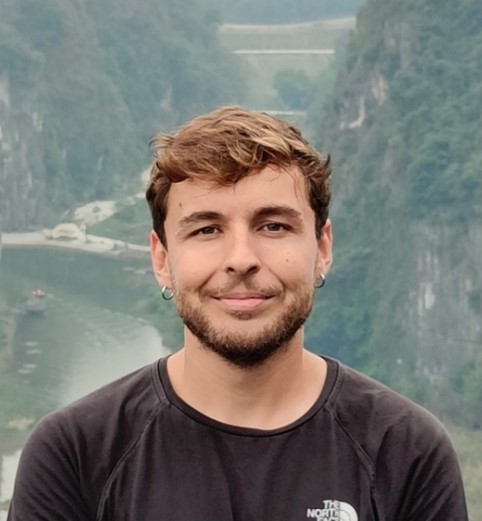}}]{Javier Pastor Galindo} is "Juan de la Cierva" Postdoctoral Researcher at the Department of Computer Systems at Universidad Politécnica de Madrid, Spain. His research focuses on combating mis/disinformation through cyber threat intelligence, simulation environments and artificial intelligence applications in both civilian and military contexts.
\end{IEEEbiography}


\begin{IEEEbiography}[{\includegraphics[width=1in,height=1.25in,clip,keepaspectratio]{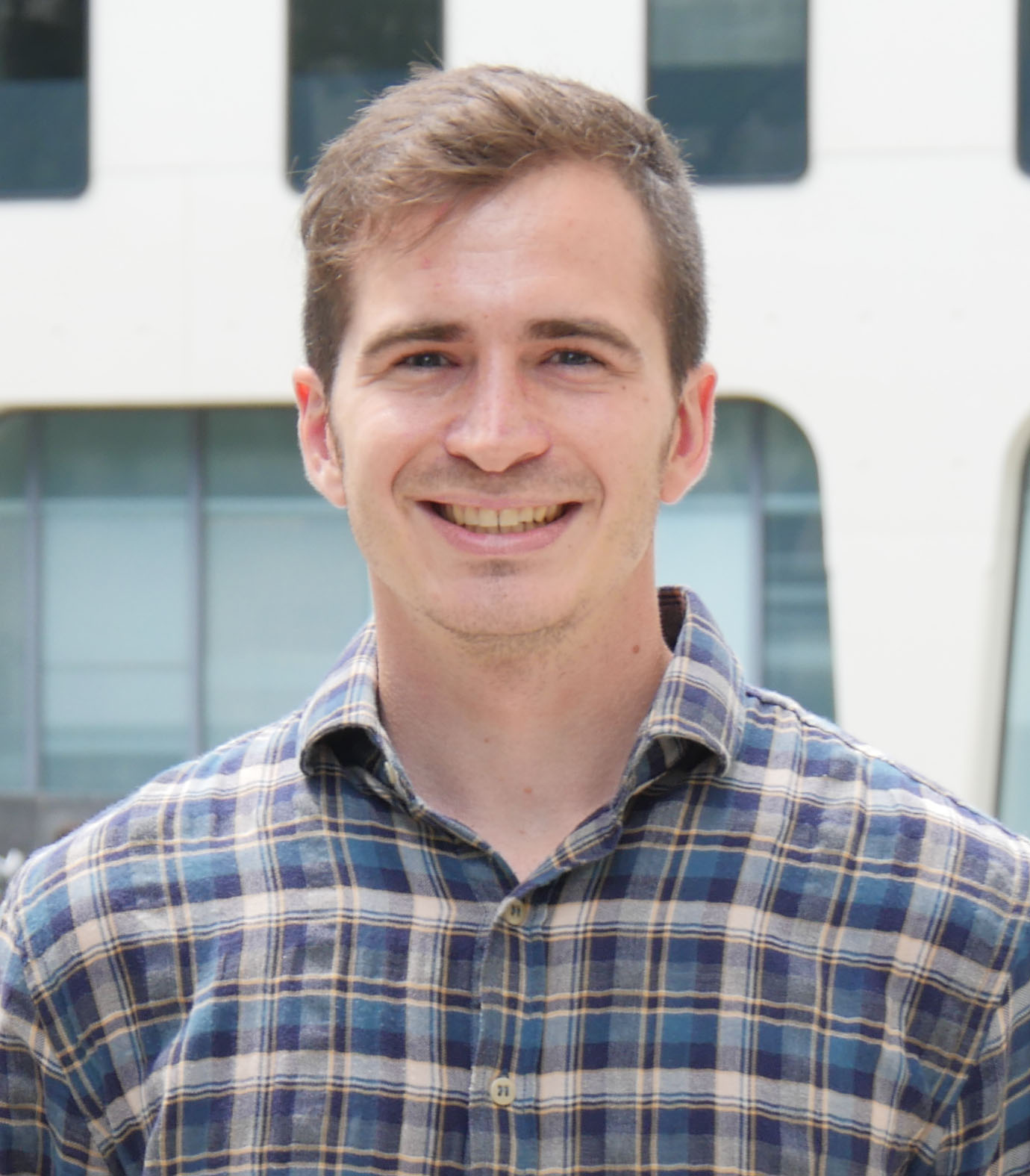}}]{Jos\'e~A.~Ruip\'erez-Valiente} (Senior Member, IEEE)
received his B.Eng. degree in telecommunications from Universidad Católica de San Antonio de Murcia in 2011 and a M.Eng. degree in telecommunications in 2013, together with his M.Sc. and Ph.D. degrees (2014 and 2017) in telematics from Universidad Carlos III of Madrid while conducting research with Institute IMDEA Networks in the area of applied data science.  He is currently an Associate Professor of Computer Science and Artificial Intelligence at the University of Murcia.
\end{IEEEbiography}




\end{document}